\documentclass[12pt]{article}

\usepackage{epsfig}
\usepackage{amssymb,amsmath}
\usepackage{multirow}
\usepackage{rotating}
\usepackage{graphicx}
\usepackage{lscape}
\usepackage{hyperref}

\newcommand{\bea}{\begin{eqnarray}}
\newcommand{\eea}{\end{eqnarray}}

 \makeatletter
\def\alt{\mathrel{\mathpalette\gl@align<}}
\def\agt{\mathrel{\mathpalette\gl@align>}}
\def\gl@align#1#2{\lower.6ex\vbox{\baselineskip\z@skip\lineskip\z@
\ialign{$\m@th#1\hfil##\hfil$\crcr#2\crcr\sim\crcr}}} \makeatother

\begin{document}

\begin{flushright}
\end{flushright}

\vspace*{1.0cm}

\begin{center}
\baselineskip 20pt 
{\Large\bf 
GUT-inspired MSSM in light of \\
muon $g-2$ and LHC results at $\sqrt{s}=13$ TeV
}
\vspace{1cm}

{\large 
Hieu Minh Tran$^{a}$\footnote{ E-mail: hieu.tranminh@hust.edu.vn} 
and 
Huong Thu Nguyen$^b$
} 
\vspace{.5cm}

{\baselineskip 20pt \it
$^a$Hanoi University of Science and Technology, 1 Dai Co Viet Road, Hanoi, Vietnam \\
\vspace{2mm}
$^b$Faculty of Physics, VNU University of Science, Vietnam National University - Hanoi, \\ 334 Nguyen Trai Road, Hanoi, Vietnam \\
\vspace{2mm}
}

\vspace{.5cm}

\vspace{1.5cm} {\bf Abstract}
\end{center}

The recent results of the LHC search for  electroweak production of supersymmetric (SUSY) particles at $\sqrt{s}=13$ TeV have shown improved lower limits for their masses.
In addition, the projected experiment E989 will be able to measure the muon anomalous magnetic moment precisely so that the experimental uncertainty can be reduced by a factor of four.
It was pointed out that if the center value of the muon $g-2$ remains unchanged the deviation between the standard model (SM) prediction and the experimental value will be as large as 7.0$\sigma$.
Such a large deviation will be solid evidence for  new physics beyond the SM.
Motivated by these results, we investigate the minimal SUSY extension of the SM with universal gaugino masses at the grand unified scale in the light of the muon $g-2$ and the updated LHC constraints. 
The squarks are assumed to be heavy and decoupled from physics at low energy scales to resemble the SM-like Higgs boson mass of 125 GeV and other bounds for squark masses at the LHC.
We have pinned down allowed windows for the lightest neutralino and the smuon masses as well as other input parameters relevant to the light SUSY sector.
The expected results of the E989 experiment play a crucial role in narrowing these windows.
The viability of the model for small mass regions can be tested at the LHC Run 3 and the High Luminosity LHC in the near future.

\thispagestyle{empty}

\newpage

\addtocounter{page}{-1}

\baselineskip 18pt

\section{Introduction}  


The standard model (SM) has successfully predicted various physical observables. 
However, there are still theoretical and experimental problems that require a careful investigation of this model. Some of them lead to the necessity of new physics beyond the SM.
It is well known that the SM suffers from the gauge hierarchy problem that originates from the existence of a scalar sector and two distant energy scales (for example, the electroweak and the Planck scales) in the theory. When taking into account quantum corrections, the quadratic divergences destabilize the electroweak scale leading to the request of fine-tuning parameters.
On the experimental side, one of the long-standing problems is the 
$3-4\sigma$ deviation between the SM prediction and the measured value of the muon anomalous magnetic dipole moment 
($a_\mu = \frac{g-2}{2}$) 
\cite{Blum:2013xva,Keshavarzi:2018mgv}.

To address these problems, physics beyond the SM is necessary. In the minimal supersymmetric (SUSY) extension of the SM (MSSM), the contributions of the superpartners of the SM particles make the theory UV-insensitive and to ameliorate the gap between the experimental value and the SM prediction of the muon $g - 2$.
Moreover, the gauge couplings in the MSSM naturally unify at a high energy scale of the order $10^{16}$ GeV, suggesting the existence of a grand unified theory (GUT).
In such a SUSY GUT scenario, the gaugino masses are interrelated at the GUT scale resulting in a specific relation among them at low energies. This property reduces the degree of freedom in the gaugino sector.

On the one hand, the measured value of the Higgs boson mass of 125 GeV \cite{Aad:2015zhl} imposes a constraint on the stop masses that are at the order of about 10 TeV 
\cite{Baer:2011ab}
assuming small stop mixing.
Due to this requirement, we assume that the squarks are all heavy and decoupled from physics at low energies in a similar way as in the split SUSY scenario 
\cite{ArkaniHamed:2004fb}.
This is in agreement with the fact that the LHC  has not found any signal for squarks \cite{Sirunyan:2018vjp}.
In the various SUSY breaking models,
correlations between the squark and slepton masses are assumed at some high energy scale
\cite{Okada:2010xe}.
Therefore, after the renormalization group (RG) running from the high scale down to the electroweak scale, the sleptons are relatively heavy 
\cite{Cao:2011sn}.
On the other hand, the muon $g-2$ constraint requires relatively light smuons and neutralinos/charginos
\cite{Belanger:2001am,Cho:2011rk}.
Such tension between the muon $g-2$ and the Higgs boson mass constraints can be resolved by a mass splitting between the stops and the smuons \cite{Ibe:2012qu}.
The tension can also be addressed with a large stop mixing parameter $X_t = A_t - \mu\cot\beta$ that increases the Higgs boson mass
\cite{Okada:2013ija}
\footnote{With large stop mixing, the stop masses can be as light as $2-4$ TeV 
\cite{Bahl:2017aev}.}.
In the meanwhile, a large value of 
$\tan\beta$ is known to enhance the muon $g-2$
\cite{Marchetti:2008hw}.
Especially, the limit 
$\tan\beta \rightarrow \infty$ was studied in \cite{Bach:2015doa}.
Other possibilities have been investigated to address this shortcoming as well
\cite{Endo:2011mc}.

Recently, the Large Hadron Collider (LHC) data at $\sqrt{s} = 13$ TeV with the luminosity of 36.1 fb$^{-1}$ have shown much improved limits on the masses of neutralinos and charginos
\cite{Aaboud:2018jiw,Aaboud:2018sua}
in comparison with the results of the previous run \cite{Aad:2014vma}.
These constraints have important consequences in the phenomenology of the MSSM \cite{Duan:2018rls} as well as other extensions \cite{Banerjee:2018eaf}.
Beside the searches at the LHC, the muon $g-2$ experiments at Fermilab (E989)
\cite{Grange:2015fou} and at J-PARC (E34) 
\cite{Mibe:2010zz} will precisely measure the muon anomalous magnetic dipole moment that will shed light on the MSSM as well as other possible new physics beyond the SM.
These two experiments are planned to reduce the experimental uncertainty of the muon $g-2$ by a factor of four 
\cite{Lusiani:2018tcd}
compared to the previous experiment E821 at Brookhaven National Laboratory (BNL)
\cite{Bennett:2002jb}.
Taking into account the recent result in theoretical evaluation of the hadronic vacuum polarization contributions to
of the muon $g-2$, the hadronic contributions to the effective QED coupling at the Z boson mass,
and the projected accuracy of the E989 experiment assuming that the center value remains intact, it was shown that the deviation between the experimental value and the SM prediction is up to $7.0\sigma$
\cite{Keshavarzi:2018mgv}.
Such a large deviation will provide solid evidence for the existence of new physics beyond the SM.

Motivated by the recent LHC limits and the projected muon g-2 experiments, we investigate a GUT-inspired MSSM where the gaugino masses satisfy a specific relation, and all the scalar superpartners are assumed to be heavy except for the sleptons.
In the paper \cite{Kowalska:2015zja}, the scenario of gaugino non-universality was studied with the LHC data at 8 TeV. The prospects at the designed center of mass energy were also examined.
Here, we will focus on the scenario of universal gaugino masses and consider the updated result from the LHC search at $\sqrt{s} = 13$ TeV, which is close to its designed energy, as well as the prospects for the muon $g-2$ experiments.
We will show that the forthcoming result of the E989 experiment at Fermilab will play a crucial role in testing the model and restricting large portions of the parameter space due to its high precision.
The paper is organized as follows.
In Section 2, the GUT-inspired MSSM is described.
In Section 3, we present the results of numerical analyses, taking into account the recent LHC constraints at $\sqrt{s}=13$ TeV, the projected precision of the muon $g-2$ experiment E989, as well as the theoretical constraint of the vacuum metastability.
Finally, Section 4 is devoted to conclusions.

\section{A GUT-inspired MSSM and the muon $g-2$}

In the MSSM, the scale dependence of the gauge couplings and the gaugino masses at one-loop are determined from their RG equations at one loop level as
\cite{Castano:1993ri}
\begin{eqnarray}
\frac{d g_i}{d t} &=&
	-\frac{1}{16\pi^2} b_i g_i^3,	\\
\frac{d M_i}{dt} &=& 
	-\frac{1}{8\pi^2} b_i g_i^2 M_i \qquad  (i=1,2,3).
\end{eqnarray}
It follows that the ratio $\frac{M_i}{g_i^2}$ is scale independence.
Here, we assume a boundary condition that all the three gauge interactions unify, and the gaugino masses are universal at the GUT scale.
Therefore, the gaugino masses satisfy the following GUT relation \cite{Martin:1997ns}:
\begin{eqnarray}
\frac{M_1}{g_1^2}
	= \frac{M_2}{g_2^2}
	= \frac{M_3}{g_3^2}
	= \frac{m_{1/2}}{g_{\text{GUT}}^2} ,
\label{GUTrelation}
\end{eqnarray}
where $m_{1/2}$ and $g_{\text{GUT}}$ are the common gaugino mass and the unified gauge coupling at the GUT scale respectively.
At low energies, the gaugino sector can be parameterized by only one parameter, the bino mass $M_1$.

In this investigation, we further assume that squarks are all heavy. Thus, the free parameters relevant to our study are the bino mass ($M_1$), the slepton soft masses ($m_{\tilde{l}_i}$), the supersymmetric Higgs mass ($\mu$) and the ratio between two vacuum expectation values ($\tan\beta$).
These assumptions is compatible with the simplified models employed by ATLAS and CMS Collaborations in analyzing data at the LHC \cite{Alwall:2008ag}. 
Here, only a few particles are relatively light and relevant for the LHC physics.
Therefore, we can straightforwardly use the limits given by the LHC searches.
When the $\mu$ parameter is large, there are only two lightest neutralinos ($\tilde{\chi}^0_{1,2}$) and one lightest chargino ($\tilde{\chi}^\pm_1$) of the gaugino/higgsino sector relevant to the searches at the LHC. 
In this case, these particles are mostly bino-like and wino-like, and their masses can be approximated by $M_1$ and $M_2 = \frac{g_2^2}{g_1^2} M_1$
\cite{Martin:1997ns}.
Since the chargino $\tilde{\chi}^\pm_1$ and the neutralino $\tilde{\chi}^0_2$ are dominated by the wino component, their interactions with the right-handed sfermions are negligible.

In this model, the SUSY particles that contribute predominantly to the anomalous muon $g-2$ are the neutralinos ($\tilde{\chi}^0_{1,2}$), the chargino ($\tilde{\chi}^\pm_1$), the smuons and the muon-sneutrino.
For convenience, we parameterize the smuon masses as
\begin{eqnarray}
m_{\tilde{\mu}_L} &=&
	M_1 + (M_2 - M_1) r ,	\\
m_{\tilde{\mu}_R} &=& m_{\tilde{\mu}_L} - \delta,
\end{eqnarray}
where the dimensionless parameter $r$ must be non-negative to avoid the situation with stable charged sleptons,
and $\delta$ is the mass splitting between right-handed and left-handed smuons.
The MSSM contribution to the muon $g-2$ is estimated at one loop level as \cite{Moroi:1995yh}
\begin{eqnarray}
\Delta a_\mu^{\text{SUSY,1}} 
	&=& \frac{\alpha m_\mu^2 \mu M_2 \tan \beta}{4\pi \sin^2 \theta_W m_{\tilde{\mu}_L}^2}
		\left[
		\frac{f_{\chi}(M_2^2 / m_{\tilde{\mu}_L}^2) - f_\chi (\mu^2/ m_{\tilde{\mu}_L}^2)}{M_2^2 -\mu^2}
		\right]		\nonumber	\\
&&	+	\frac{\alpha m_\mu^2 \mu M_1 \tan \beta}{4\pi \cos^2 \theta_W (m_{\tilde{\mu}_R}^2 - m_{\tilde{\mu}_L}^2 )}
		\left[
		\frac{f_N(M_1^2 / m_{\tilde{\mu}_R}^2)}{m_{\tilde{\mu}_R}^2}
		 - \frac{f_N (M_1^2/ m_{\tilde{\mu}_L}^2)}{m_{\tilde{\mu}_L}^2}
		\right] ,
\end{eqnarray}
where the loop functions are defined as
\begin{eqnarray}
f_\chi (x)	&=&
	\frac{x^2 - 4x +3 + 2\ln x}{(1-x)^3} \,  , 	\\
f_N	(x)		&=&	
	\frac{x^2 - 1 - 2x \ln x}{(1-x)^3} \, .
\end{eqnarray}
where $m_\mu$ and $\theta_W$ are the muon mass and the Weinberg mixing angle respectively.
The SUSY contribution to the muon $g-2$, taking into account two loop diagrams, is determined as \cite{Endo:2013lva}
\begin{eqnarray}
\Delta a_\mu^{\text{SUSY}} &=& 
	\frac{1}{1 + \Delta_\mu}
	\left(
	1  - \frac{4\alpha}{\pi} \ln \frac{M_1}{m_\mu}
	\right)
	\left[
	1 + 
	\frac{2 \alpha \Delta b}{4\pi}
		\ln \frac{M_\text{soft}}{M_1}
	  + 
	\frac{1}{4\pi} \frac{9}{4}\alpha_2
		\ln \frac{M_2}{M_1}
	\right] 
	\Delta a_\mu^{\text{SUSY,1}} \, .
\label{amu}
\end{eqnarray}
The first factor in the above equation comes from the correction to the muon Yukawa coupling constant with $\Delta_\mu$ given as \cite{Marchetti:2008hw}
\begin{eqnarray}
\Delta_\mu &=&
	- \mu \tan\beta \frac{g_2^2 M_2}{16\pi^2} 
	\left[	
	f(m_1^2, m_2^2, m_{\tilde{\nu}_\mu}^2) +
	\frac{1}{2}
	f(m_1^2, m_2^2, m_{\tilde{\mu}_L}^2)
	\right]		\nonumber	\\
&&	- \mu \tan\beta \frac{g_1^2 M_1}{16\pi^2} 
	\left[	
	f(\mu^2, M_1^2, m_{\tilde{\nu}_R}^2) -
	\frac{1}{2}
	f(\mu^2, M_1^2, m_{\tilde{\nu}_L}^2) -
	f(M_1^2, m_{\tilde{\nu}_L}^2, m_{\tilde{\nu}_R}^2)
	\right] \,	,	\label{Delta}
\end{eqnarray}
where the loop function is defined as
\begin{eqnarray}
f(a,b,c) &=& 
	- \frac{ab \ln \frac{a}{b} + 
			bc \ln \frac{b}{c} +
			ca \ln \frac{c}{a} }
	{(a-b)(b-c)(c-a)} \, ,
\end{eqnarray}
and the chargino masses, $m_{1,2}$, are given as
\begin{eqnarray}
m_{1,2}^2 &=&
	\frac{1}{2}
	\left[
	(M_2^2 + \mu^2 + 2M_W^2) \mp
	\sqrt{ (M_2^2 + \mu^2 + 2M_W^2)^2 - 
		4 (M_2 \mu - M_W^2 \sin 2\beta)^2
	}
	\right] \, .
\end{eqnarray}
This factor turns out to be important for the cases of large $\tan\beta$.
The second factor written inside the round brackets of Eq. (\ref{amu}) is due to the QED corrections to the muon $g-2$ 
\cite{Degrassi:1998es}.
In the region of the parameter space that we are considering, the magnitude of this two-loop contribution is about $\mathcal{O}(10\%)$ of the one-loop contribution.
It was found that the non-logarithmic terms  of the QED corrections are negligibly small in comparison to the leading logarithmic one \cite{vonWeitershausen:2010zr}.
The third factor in the right hand side of Eq. (\ref{amu}) is caused by the corrections to the bino-smuon-muon coupling.
The term with the coefficient $\Delta b$ originates from the contributions of the SM particles and the light sparticles when the heavy sparticles are decoupled at the scale $M_{\text{soft}} \sim \mathcal{O}(10)$ TeV.
Hence, in our case we have $\Delta b = \frac{41}{6} - n_{\tilde{l}}$ where $n_{\tilde{l}}$ is the number of light slepton generations involve in the correction.
Note that $n_{\tilde{l}} =3 $ when all three generations of sleptons are light, and
$n_{\tilde{l}} =2$ in the case that the staus are heavy and decoupled.
The term of $\alpha_2$ is the correction related to the wino contribution that is small compared to the above contributions.

The anomalous muon $g-2$ is enhanced by  the large left-right mixing that is controlled by $\mu \tan\beta$.
However, a too large value of $\mu \tan\beta$ might lead to a charged breaking vacuum that is deeper than the electroweak breaking vacuum 
\cite{Endo:2013lva, Kitahara:2013lfa}.
In such case, the phenomenological consistency required that the lifetime of the electroweak breaking vacuum 
\cite{Coleman:1977py}
 should be longer than the age of the universe.
Thus, the value of $\mu \tan \beta$ is subjected to the condition of vacuum metastability given by the following fitting formula 
\cite{Endo:2013lva}
\begin{eqnarray}
\left| m^2_{\tilde{l}_{LR}}
\right|	&\leq &
	\eta_l
	\left[
	1.01 \times 10^{2} \text{ GeV}
		\left(
		\sqrt{m_{\tilde{l}_L} m_{\tilde{l}_R}} +
		m_{\tilde{l}_L} + 1.03 m_{\tilde{l}_R}
		\right)
	- 2.27 \times 10^4 \text{ GeV}^2
	\right. \nonumber \\
&&	\qquad
	\left.
	+ \frac{2.97 \times 10^6 \text{ GeV}^3}{m_{\tilde{l}_L} + m_{\tilde{l}_R}}
	- 1.14 \times 10^8 \text{ GeV}^4
		\left(
		\frac{1}{m^2_{\tilde{l}_L}} +
		\frac{0.983}{m^2_{\tilde{l}_L}}
		\right)
	\right] ,	\label{vacuum}
\end{eqnarray}
where $\eta_l \sim \mathcal{O}(1)$, and the left-right mixing mass of slepton is determined as
\begin{eqnarray}
m^2_{\tilde{l}_{LR}} &=& 
-\frac{m_l}{1 + \Delta_l} \mu \tan\beta.
\end{eqnarray}
Here, $\Delta_l$ determined by a formula similar to Eq. (\ref{Delta}) is a correction to the lepton Yukawa coupling.
Since the left hand side of Eq. (\ref{vacuum}) is proportional to the lepton masses, the condition is most severe for stau in the case that all the three slepton masses are degenerate.
In the case that the masses of the first two generation sleptons are degenerate and the staus are decoupled (for example, in the split-family scenario), the condition (\ref{vacuum}) is more severe for the smuon.
%
%
%
%
%
%
Another issue related to the large left-right mixing is the possible blow-up of the down-type Yukawa couplings when $\tan\beta$ becomes too large. 
According to the papers \cite{Bach:2015doa} and \cite{Brignole:2002bz}, the parameter regions that we consider in the next section are safe in terms of a perturbation theory.

\section{Numerical analysis
}



With the measured value of the muon anomalous magnetic moment at the E821 experiment \cite{Bennett:2002jb} carried out at BNL, the world average of the muon $g-2$ becomes \cite{Tanabashi:2018oca}
\begin{eqnarray}
a_\mu^\text{exp} = 11 659 209.1 (5.4) (3.3) \times 10^{-10}.	\label{e821}
\end{eqnarray}
Employing a new data combination method in a bias free approach, the recent estimation for the SM prediction of the muon $g-2$ reads
\cite{Keshavarzi:2018mgv}
\begin{eqnarray}
a_\mu^\text{SM} 
	= a_\mu^\text{QED} + a_\mu^\text{EW} + a_\mu^\text{had}
	= (11 659 182.04 \pm 3.56) \times 10^{-10},	\label{SMprediction}
\end{eqnarray}
where 
$a_\mu^\text{QED}$, 
$a_\mu^\text{EW}$, and 
$a_\mu^\text{had}$
are respectively
the quantum electrodynamics (QED) contribution, 
the electroweak contribution, and 
the hadronic contributions due to the vacuum polarization and the light-by-light scattering.
The uncertainty of $a_\mu^\text{QED}$ is due to the uncertainties on 
the lepton masses, 
the 4-loop contributions, 
the 5-loop contributions, and 
the determination of the fine structure constant $\alpha$ \cite{Aoyama:2012wk}.
The uncertainty of $a_\mu^\text{EW}$ includes those on 
the unknown 3-loop contributions, 
the neglected 2-loop terms, 
the hadronic loops, and 
the measured masses of the involved SM particles \cite{Gnendiger:2013pva}.
The dominant sources of the theoretical uncertainty on the SM prediction are those of the hadronic vacuum polarization and the hadronic light-by-light scattering contributions ($a_\mu^\text{had}$) 
which originate from 
the statistical and systematic errors of experimental data, and 
the uncertainties of the vacuum polarization corrections and the final state radiative corrections to the cross section data \cite{Keshavarzi:2018mgv}.
Accordingly, the discrepancy between the experimental value and the SM prediction is 3.7 standard deviation:
\begin{eqnarray}
\Delta a_\mu^\text{exp}  = a_\mu^\text{SM} - a_\mu^\text{exp} 
= (27.06 \pm 7.26) \times 10^{-10}.	\label{da_current}
\end{eqnarray}
The experiment E989 at Fermilab  \cite{Grange:2015fou} is designed to collect 21 times the E821 data set with improved instrumentation.
Thus, this experiment is expected to achieve a better uncertainty about four times smaller than that of the experiment E821. 
Assuming the same center value of $\Delta a_\mu^\text{exp}$ as in the current result (\ref{da_current}), the projected deviation between the SM prediction and the experimental value is
\begin{eqnarray}
\Delta a_\mu^\text{exp} = (27.06 \pm 3.87) \times 10^{-10},	\label{da_projected}
\end{eqnarray}
corresponding to a 7.0$\sigma$ discrepancy \cite{Keshavarzi:2018mgv}.


Beside the muon $g-2$ constraint, the light superpartners are also subjected to the limits obtained from the LHC data.
We consider the constraints from the LHC searches for light sleptons, neutralinos and charginos using 36.1 fb$^{-1}$ of proton-proton collision data at the center of mass energy $\sqrt{s} = 13$ TeV \cite{Aaboud:2018jiw, Aaboud:2018sua}.
In this model, the neutralino $\tilde{\chi}_2^0$ and the chargino $\tilde{\chi}_1^\pm$ cannot decay into squarks due to their heavy masses.
The decay products of $\tilde{\chi}_2^0$ and $\tilde{\chi}_1^\pm$ depend on the relation between their masses and the slepton ones.
In any case, the final production of these superparticles always includes the lightest neutralino $\tilde{\chi}^0_1$ that is stable under the R-parity conservation.


\textbf{\underline{\textit{Case 1}:}} 
$M_1 < m_{\tilde{l}_i} < M_2$. Equivalently, the range of the parameter $r$ is $0 < r < 1$.

In this case, on the one hand, the second lightest neutralino $\tilde{\chi}_2^0$ once produced can decay into a lepton and a same flavor slepton that finally decays into its SM partner and the lightest neutralino:
$\tilde{\chi}_2^0 \rightarrow \tilde{l}^\mp (\tilde{\nu}) + l^\pm (\nu)
		\rightarrow \tilde{\chi}_1^0 + l^\mp (\nu) + l^\pm (\nu)$.
On the other hand, to end up with the same final states, the neutralino $\tilde{\chi}_2^0$ can also decay directly to the lightest neutralino $\tilde{\chi}_1^0$ and a $Z^0/h$ boson that subsequently decays into a pair of leptons:
$\tilde{\chi}_2^0 \rightarrow \tilde{\chi}_1^0 + Z^0/h
		\rightarrow \tilde{\chi}_1^0 + l^- (\nu) + l^+ (\nu)$.
Similarly, the light chargino $\tilde{\chi}_1^\pm$ once produced can decay to a charged slepton and a neutrino of the same generation, or a charged lepton and a sneutrino of the same generation.
The slepton/sneutrino then decays into the lightest neutralino $\tilde{\chi}_1^0$ and the corresponding SM partner:
$\tilde{\chi}_1^\pm \rightarrow \tilde{l}^\pm (\tilde{\nu}_l) + \nu_l (l^\pm)
			\rightarrow \tilde{\chi}_1^0 + l^\pm + \nu_l$.
The charginos $\tilde{\chi}_1^\pm$ can also decay into the lightest neutralino and leptons via the mediation of a $W$ boson:
$\tilde{\chi}_1^\pm \rightarrow \tilde{\chi}_1^0 + W^\pm
			\rightarrow \tilde{\chi}_1^0 + l^\pm + \nu_l$.

The LHC searches for the chargino and neutralino productions have shown the limits for the masses of these particles.
The current best limits are obtained from the analysis of the data at the center of mass energy $\sqrt{s}=13$ TeV with the luminosity of 36.1 fb$^{-1}$
\cite{Aaboud:2018jiw,Aaboud:2018sua}.
The search for the $\tilde{\chi}_1^+ \tilde{\chi}_1^-$ pair production with $\tilde{l}$ mediated decays was analyzed by the ATLAS Collaboration in the 2 leptons and 0 jet channel \cite{Aaboud:2018jiw}.
For the GUT-inspired model, we can extract the limit for the bino mass as%
\begin{eqnarray}
M_1 \gtrsim 350 \text{ GeV}  \qquad (95\% \, \text{C.L.}).	\label{8a}
\end{eqnarray}
The search for the associated $\tilde{\chi}_1^\pm \tilde{\chi}_2^0$ production with $\tilde{l}$ mediated decays in the 3 leptons channel imposes even more severe constraint \cite{Aaboud:2018jiw}.
In the GUT-inspired model, the result implies that
\begin{eqnarray}
M_1 \gtrsim 700 \text{ GeV}  \qquad (95\% \, \text{C.L.}).	\label{8c}
\end{eqnarray}
It is worth noting that the constraints (\ref{8a}) and (\ref{8c}) are applied when the slepton masses are assumed to be midway between bino and wino masses, namely
$m_{\tilde{l}_L}=\frac{M_2 - M_1}{2}$.

The $\tilde{l} \tilde{l}$ production has been analyzed in the 2 leptons plus 0 jet channel \cite{Aaboud:2018jiw}.
Here, the slepton $\tilde{l}$ decays directly into a same flavor lepton $l$ and the lightest neutralino $\tilde{\chi}_1^0$ with 100\% branching ratio:
$\tilde{l}^\pm \tilde{l}^\mp \rightarrow
	l^\pm + l^\mp + 2 \tilde{\chi}^0_1$.
In the considered model, the limits for slepton and the neutralino masses read as
\begin{eqnarray}
M_1 \gtrsim 300 \text{ GeV}  \qquad (95\% \, \text{C.L.}),	\label{8b1}\\
m_{\tilde{l}} \gtrsim 510 \text{ GeV}  \qquad (95\% \, \text{C.L.})	\label{8b2}.
\end{eqnarray}


\textbf{\underline{\textit{Case 2}:}} 
$M_1 < M_2 \leq m_{\tilde{l}_i}$. Equivalently, the range of the parameter $r$ is $r \geq 1$.

In this case, the neutralino $\tilde{\chi}_2^0$ and the chargino $\tilde{\chi}_1^\pm$ can respectively decay into $Z^0/h$ and $W^\pm$ bosons and the lightest neutralino $\tilde{\chi}^0_1$.
The search for the associated $\tilde{\chi}_1^\pm \tilde{\chi}_2^0$ production with $W/Z$-mediated decays has been carried out in the 2 leptons + jets channel and the 3 leptons channel
\cite{Aaboud:2018jiw,Aaboud:2018sua}.
The combined limit for the bino mass extracted in this model is as follows
\begin{eqnarray}
M_1 \gtrsim 300 \text{ GeV}  \qquad (95\% \, \text{C.L.}).	\label{8d13d}
\end{eqnarray}


Given the GUT relation for gaugino masses in the model (\ref{GUTrelation}), the above constraints for the bino mass imply the lower limit for the gluino mass as
\begin{eqnarray}
M_3 \gtrsim 2 \text{ TeV} . 	
\end{eqnarray}
With the squark mass scale of $\mathcal{O}(10)$ TeV, the gluinos hadronize before decaying \cite{Kilian:2004uj}.
This indirect bound for the gluino mass in the model is consistent with the current LHC limit \cite{Aaboud:2017vwy}. 



\begin{figure}[h]
\centering
\includegraphics[width=15cm]{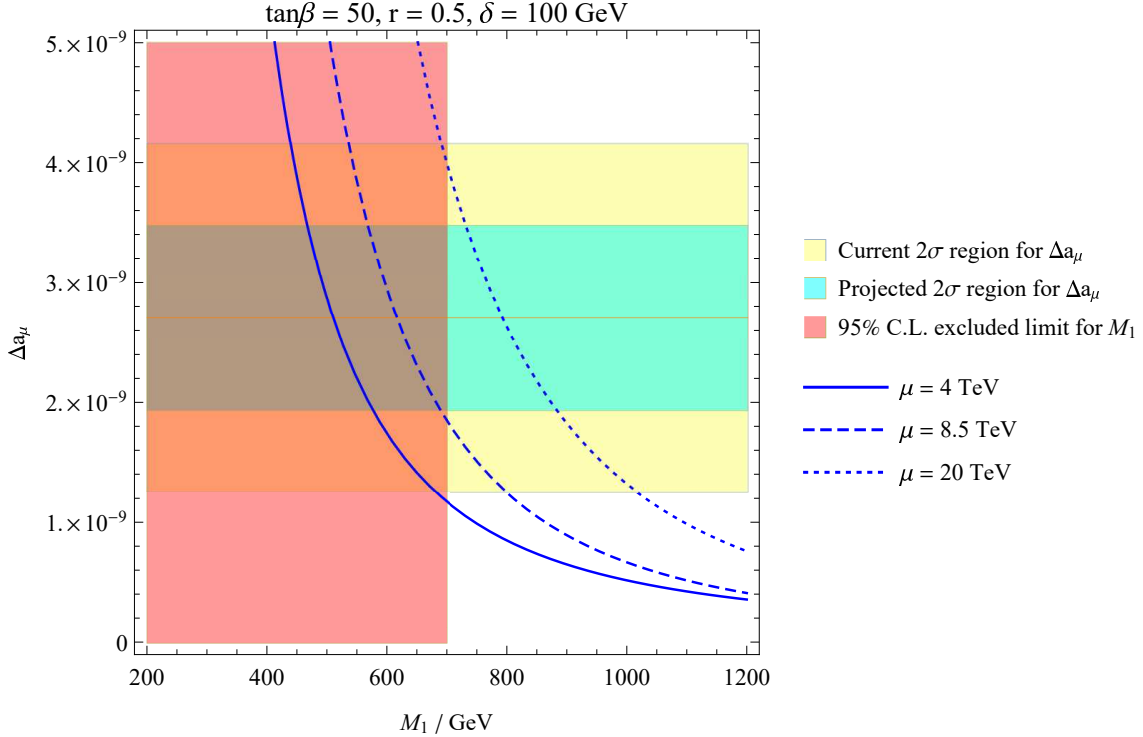}
\caption{
\label{fig1}
The SUSY contribution to the muon anomalous magnetic moment ($\Delta a_\mu$) as a function of the bino mass $M_1$ in the case 
$\tan\beta = 50$, 
$r = 0.5$,
$\delta = 100$ GeV, 
and $\mu = 4, 8.5$, and 20 TeV.
}
\end{figure}
\begin{figure}[h]
\centering
\includegraphics[width=10cm]{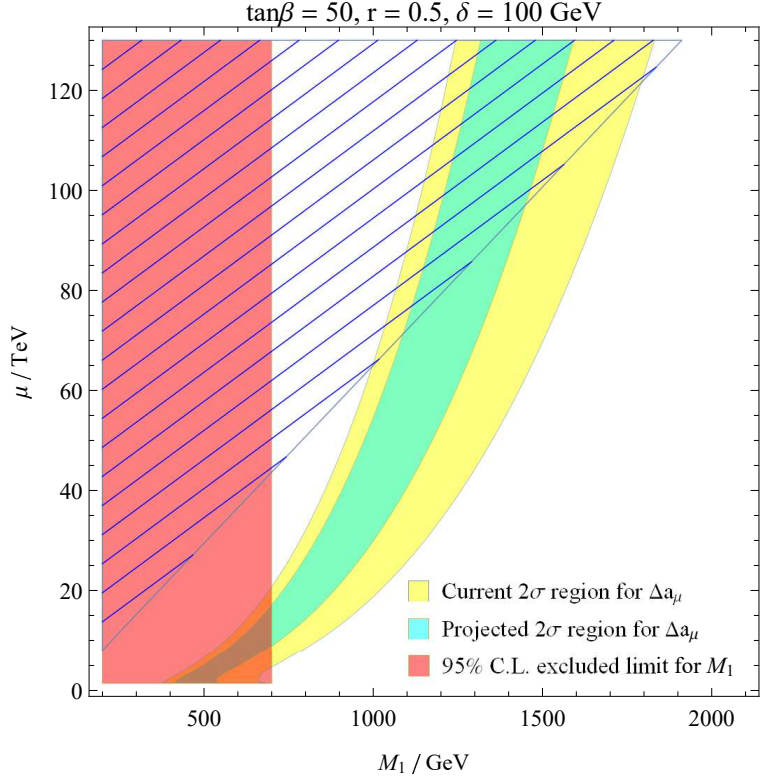}
\caption{
\label{fig2}
The phenomenological constraints on the plane of $(M_1, \mu)$. The other input parameters are fixed as
$\tan\beta = 50$, 
$r = 0.5$,
$\delta = 100$ GeV.
The hatched region is excluded by the vacuum metastability condition.
}
\end{figure}

In the numerical analysis, we investigate the impact of the current and projected muon $g-2$ constraints, the LHC limits on the bino and slepton masses, and the constraint from the vacuum metastability condition on the parameter space of the model.
Here, all the squarks are heavy, and their mass scale is taken to be $m_{\tilde{q}} = 10$ TeV.
The staus are also assumed to be heavy enough to decouple from physics at the LHC.
For the case 1, firstly, we consider the special case in which the left handed smuon masses are in the middle between the wino and bino masses, $m_{\tilde{\mu}_L} = \dfrac{M_2 - M_1}{2}$.
In Figure \ref{fig1}, we show the dependence of the SUSY contribution to the anomalous muon $g-2$ where the inputs  are set as 
$\tan\beta = 50$, 
$r = 0.5$,
$\delta = 100$ GeV.
The solid, dash and dotted curves correspond to three representative values (4, 8.5, 20 TeV) of the $\mu$ parameter.
The 95\% C.L. excluded region from the LHC search for the associated 
$\tilde{\chi}^\pm_1 \tilde{\chi}^0_2$ production in the 3 leptons channel (\ref{8c}) is shown in the red color.
The yellow and cyan bands correspond to the current and projected $2\sigma$ regions for $\Delta a_\mu$.
As expected, $\Delta a_\mu$ is enhanced for smaller values of $M_1$ and larger values of $\mu$.
We see that the lower limit on the bino mass (\ref{8c}) has ruled out the scenarios with $\mu \lesssim 4$ TeV.
In the near future, the projected result (\ref{da_projected}) can exclude the values of the $\mu$ parameter up to 8.5 TeV.
In Figure \ref{fig2}, the excluded region by the LHC search, and the allowed regions by the muon $g-2$ experiments are depicted with the same colors in the parameter space of ($M_1, \mu$).
The parameter points in the hatched region lead to the unstable electroweak vacuum with the lifetime shorter than the age of the universe. 
Therefore, this region is ruled out by the constraint (\ref{vacuum}).
The combination of the vacuum metastability  condition (\ref{vacuum}) and the current muon $g-2$ constraint (\ref{da_current}) results in the upper limits for the bino mass and the SUSY Higgs mass:
\begin{eqnarray}
M_1 & \lesssim & 1.8 \text{ TeV,}	\\
\mu & \lesssim & 120 \text{ TeV.}
\end{eqnarray}
Taking into account the LHC limit (\ref{8c}) and the projected muon $g-2$ constraint (\ref{da_projected}), the windows for these two parameters are given by
\begin{eqnarray}
700 \text{ GeV} & \lesssim \, M_1 \,  \lesssim & 1450 \text{ GeV,}	\\
8.5 \text{ TeV} & \lesssim \, \mu \, \lesssim & 97 \text{ TeV.}
\end{eqnarray}

\begin{figure}[h]
\centering
\includegraphics[width=10cm]{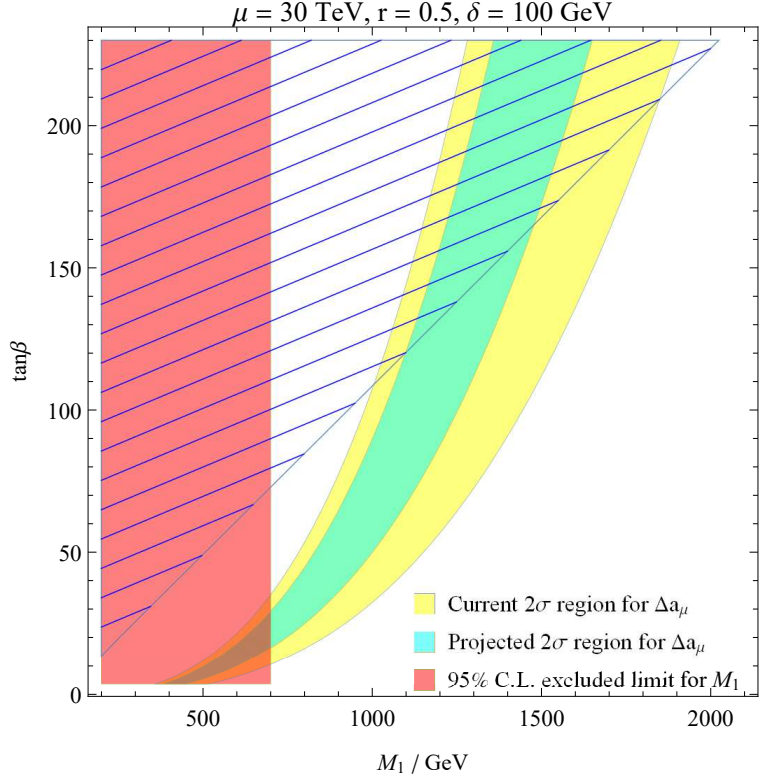}
\caption{
\label{fig3}
The phenomenological constraints on the plane of $(M_1, \tan\beta)$. 
The other input parameters are fixed as
$\mu = 30$ TeV, 
$r = 0.5$,
$\delta = 100$ GeV.
The hatched region is excluded by the vacuum metastability condition.
}
\end{figure}

\begin{figure}[h]
\centering
\includegraphics[width=10cm]{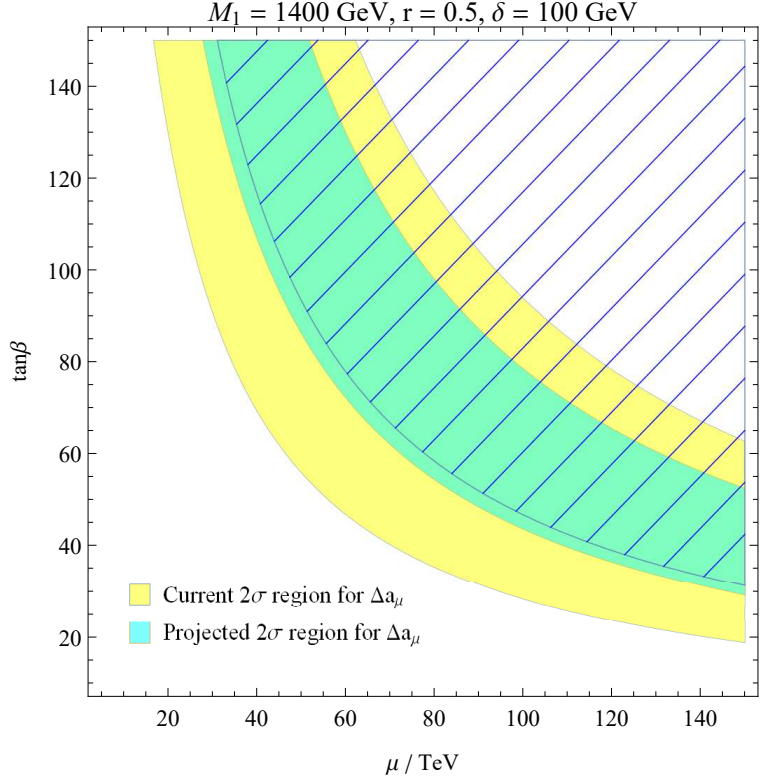}
\caption{
\label{fig4}
The phenomenological constraints on the plane of $(\mu, \tan\beta)$. 
The other input parameters are fixed as
$M_1 = 1400$ GeV, 
$r = 0.5$,
$\delta = 100$ GeV.
The hatched region is excluded by the vacuum metastability condition.
}
\end{figure}

In Figure \ref{fig3}, the constrained regions are plotted in the plane of $(M_1, \tan\beta)$.
Here, we fix the values of other input parameters as 
$\mu = 30$ TeV,
$r = 0.5$,
$\delta = 100$ GeV.
The combination of the LHC limit for the bino mass (\ref{8c}) and the muon $g-2$ constraint results in a lower bound for $\tan\beta$.
In the meanwhile the vacuum metastability condition (\ref{vacuum}) and the muon $g-2$ constraint imply an upper bound for $\tan\beta$.
For the current $2\sigma$ limits for the muon $g-2$, these bounds specify the allowed region for $\tan\beta$:
\begin{eqnarray}
10 & \lesssim \, \tan\beta \, \lesssim & 210.
\end{eqnarray}
For the projected $2\sigma$ limits for the muon $g-2$, the window for $\tan\beta$ will be narrowed
\begin{eqnarray}
15 & \lesssim \, \tan\beta \, \lesssim & 165.
\end{eqnarray}
From Figures \ref{fig2} and \ref{fig3}, we see that for a larger value of the bino mass $M_1$, larger values of $\mu$ and/or $\tan\beta$ are required to maintain an adequate SUSY contribution to the muon $g-2$ to explain the experimental result.
In Figure \ref{fig4}, the constraints are shown in the plane of ($\mu, \tan\beta$) for fixed values of the other parameters:
$M_1 =1400$ GeV,
$r = 0.5$, 
$\delta = 100$ GeV.
For a given value of the muon $g-2$, there is a dependence of $\tan\beta$ on the  $\mu$ parameter that can be approximated by  a hyperbolic function.
With this particular choice of the inputs, although the LHC searches for the neutralino and chargino productions become challenging \cite{Kowalska:2015zja},
the projected muon $g-2$ limits and the vacuum metastability condition play an important role to rule out the parameter space.
Here, only a tiny strip of the parameter space is allowed after taking these two constraints into consideration.
For $M_1 \gtrsim 1500$ GeV, the projected muon $g-2$ result from the E989 experiment 
requires too large $\mu \tan\beta$ such that it becomes inconsistent with the metastability condition of the electroweak breaking vacuum. 
Therefore, the whole parameter space for this case will be excluded if the center value of $\Delta a_\mu$ is unchanged.

\begin{figure}[h]
\centering
\includegraphics[width=15cm]{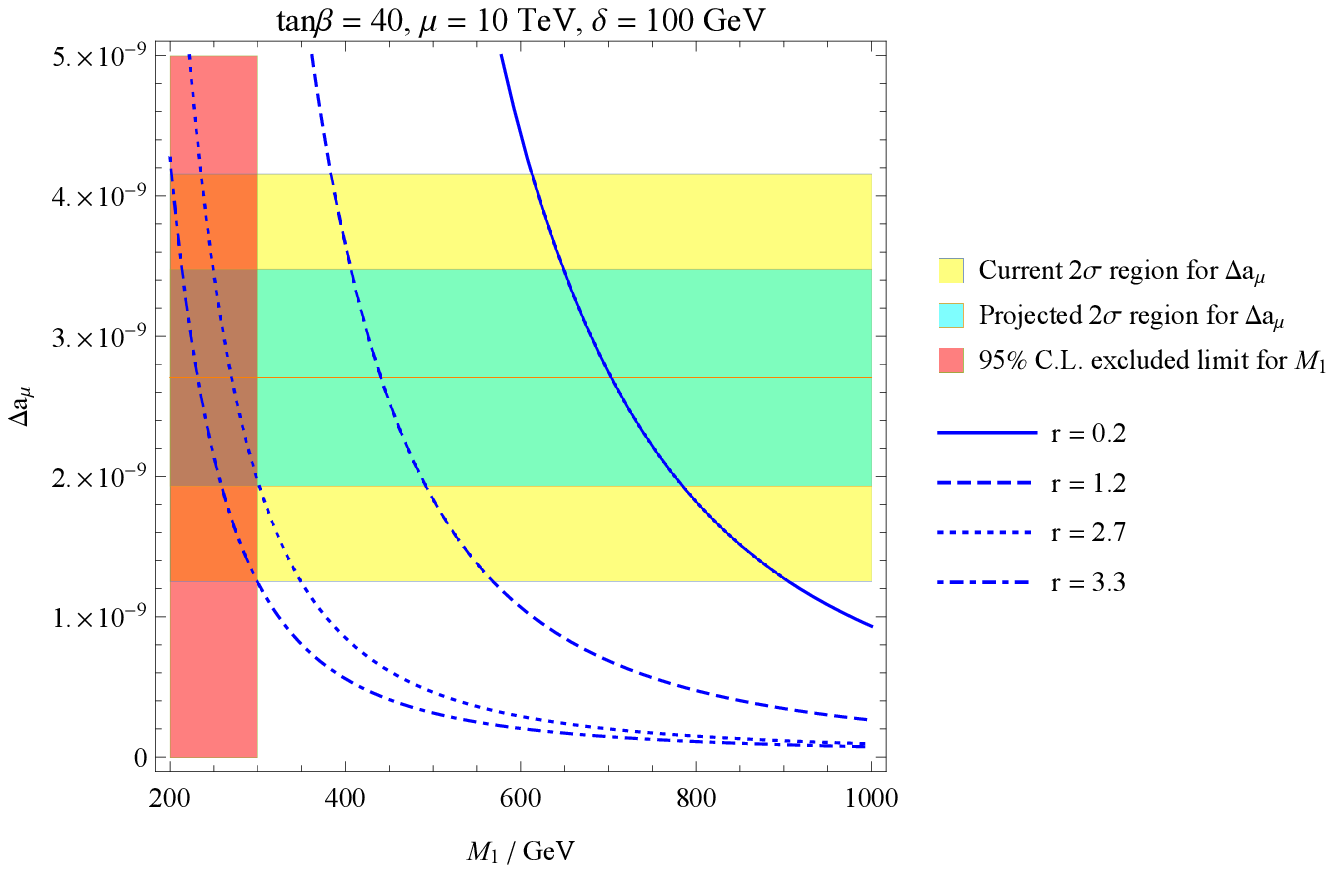}
\caption{
\label{fig5}
The SUSY contribution to the muon anomalous magnetic moment ($\Delta a_\mu$), as a function of the bino mass $M_1$ in the case $\tan\beta = 40$,
$\mu = 10$ TeV,
$\delta = 100$ GeV.
}
\end{figure}

When the specific relation between the slepton mass and the gaugino masses is relaxed 
(namely, $r \neq 0.5$), the lower bound on $M_1$ is not as severe as that in Eq. (\ref{8c}).
Instead, we employ the lower bound given by Eqs. (\ref{8b1}) and (\ref{8d13d}).
The SUSY contribution to the muon $g-2$ is depicted in Figure \ref{fig5} as a function of the bino mass $M_1$.
In this figure, the other parameters are set as 
$\tan\beta = 40$,
$\mu = 10$ TeV,
$\delta = 100$ GeV.
The solid, dash, dotted and dash-dotted curves corresponding to the cases $r =$ 0.2, 1.2, 2.7, and 3.3 respectively.
The color convention is similar to that of Figure \ref{fig1}.
In order to satisfy the muon $g-2$ constraint, the case with a larger value of $r$ requires a smaller value of $M_1$.
It is due to the fact that with increasing smuon masses, the contribution to the muon $g-2$ from the smuons is reduced.
To compensate such a reduction, lighter neutralinos and charginos will enhance the total SUSY loop contribution to $\Delta a_\mu$.
The combination of the LHC lower limit on $M_1$ and the current muon $g-2$ constraint rules out the cases with $r \gtrsim 3.3$.
Taking into account the projected E989 experimental result, upper bound on the parameter $r$ can be reduced to 2.7.

\begin{figure}[h]
\centering
\includegraphics[width=10cm]{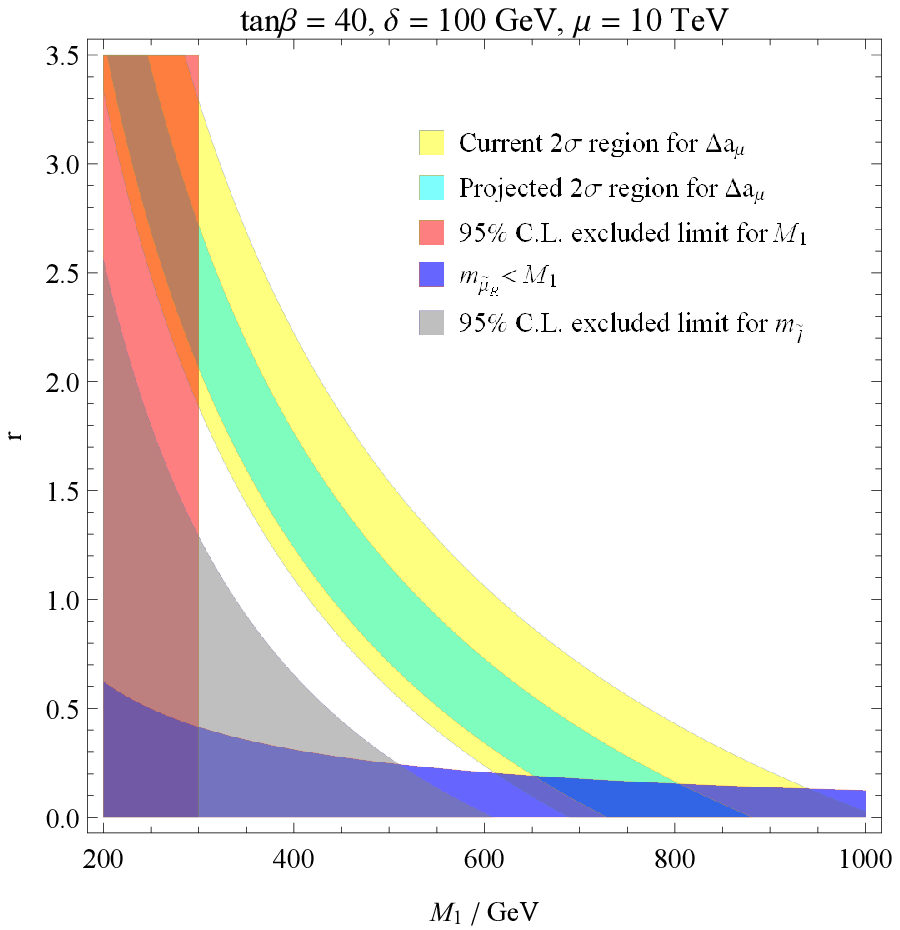}
\caption{
\label{fig6}
The phenomenological constraints in the plane of $(M_1, r)$.
The other input parameters are fixed as
$\tan\beta = 40$,
$\delta = 100$ GeV,
$\mu = 10$ TeV.
}
\end{figure}

\begin{figure}[h]
\centering
\includegraphics[width=10cm]{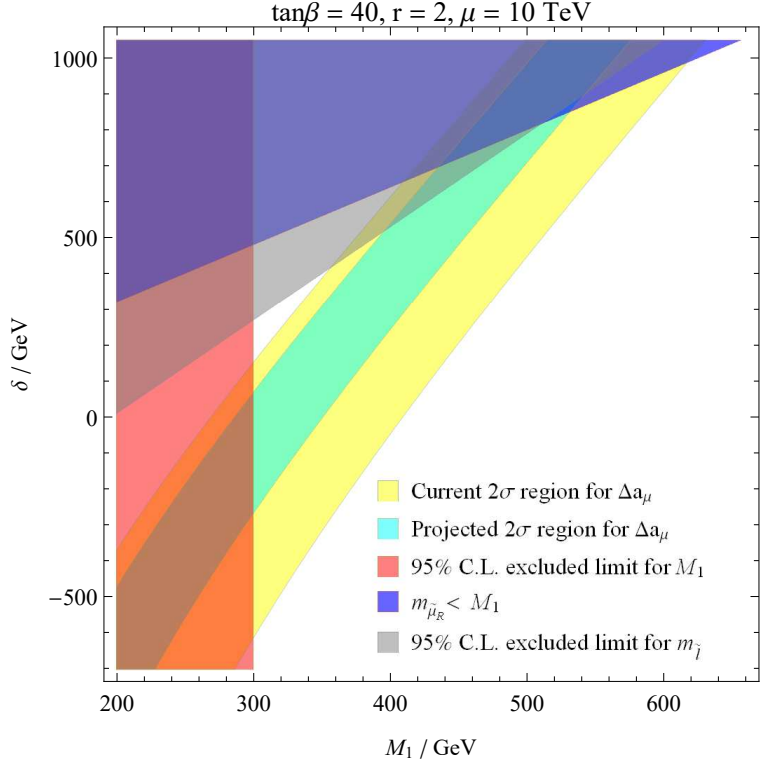}
\caption{
\label{fig7}
The phenomenological constraints in the plane of $(M_1, \delta)$.
The other input parameters are fixed as
$\tan\beta = 40$,
$r = 2$,
$\mu = 10$ TeV.
}
\end{figure}

In Figure \ref{fig6}, the constrained regions are plotted in the plane of ($M_1, r$) for fixed values of the other parameters:
$\tan\beta = 40$,
$\delta = 100$ GeV,
$\mu = 10$ TeV.
For this parameter setting, the 95\% C.L. excluded limits for $m_{\tilde{\mu}}$ is consistent with the muon $g-2$ constraint.
In addition to the previously considered constraints, we require that the right-handed smuon must be heavier than the lightest neutralino to avoid a stable charged slepton.
Therefore, the blue region in this figure is excluded.
With the current constraints, we find the window for the parameters as follows
\begin{eqnarray}
300 & \lesssim \, M_1 \, \lesssim & 950 \text{ (GeV)},	\\
0.1 & \lesssim \, r \, \lesssim & 3.3.
\end{eqnarray}
Considering the projected muon $g-2$ result, these windows become narrower
\begin{eqnarray}
300 & \lesssim \, M_1 \, \lesssim & 800 \text{ (GeV)},	\\
0.15 & \lesssim \, r \, \lesssim & 2.75.
\end{eqnarray}
In Figure \ref{fig7}, we fix the value of $r$ parameter to be 2, while relaxing the mass splitting $\delta$.
For too large mass splitting between left-handed and right-handed smuons, the right-handed one could become lighter than the neutralino $\tilde{\chi}^0_1$. Hence, this blue region is excluded.
The corresponding windows for $M_1$ and $\delta$ in this case are
\begin{eqnarray}
300 & \lesssim \, M_1 \, \lesssim & 620 \text{ (GeV)},	\\
-600 & \lesssim \, \delta \, \lesssim & 1000 \text{ (GeV)}, 
\end{eqnarray}
for the current limits from the LHC data and the muon $g-2$.
With the projected bounds on the muon $g-2$, these windows become smaller:
\begin{eqnarray}
300 & \lesssim \, M_1 \, \lesssim & 530 \text{ (GeV)},	\\
-250 & \lesssim \, \delta \, \lesssim & 850 \text{ (GeV)}. 
\end{eqnarray}
We observe that the constraint on the slepton masses (\ref{8b2}) rules out a part of the parameter region that is in agreement with the muon $g-2$ measurement.

\begin{figure}[h]
\centering
\includegraphics[width=10cm]{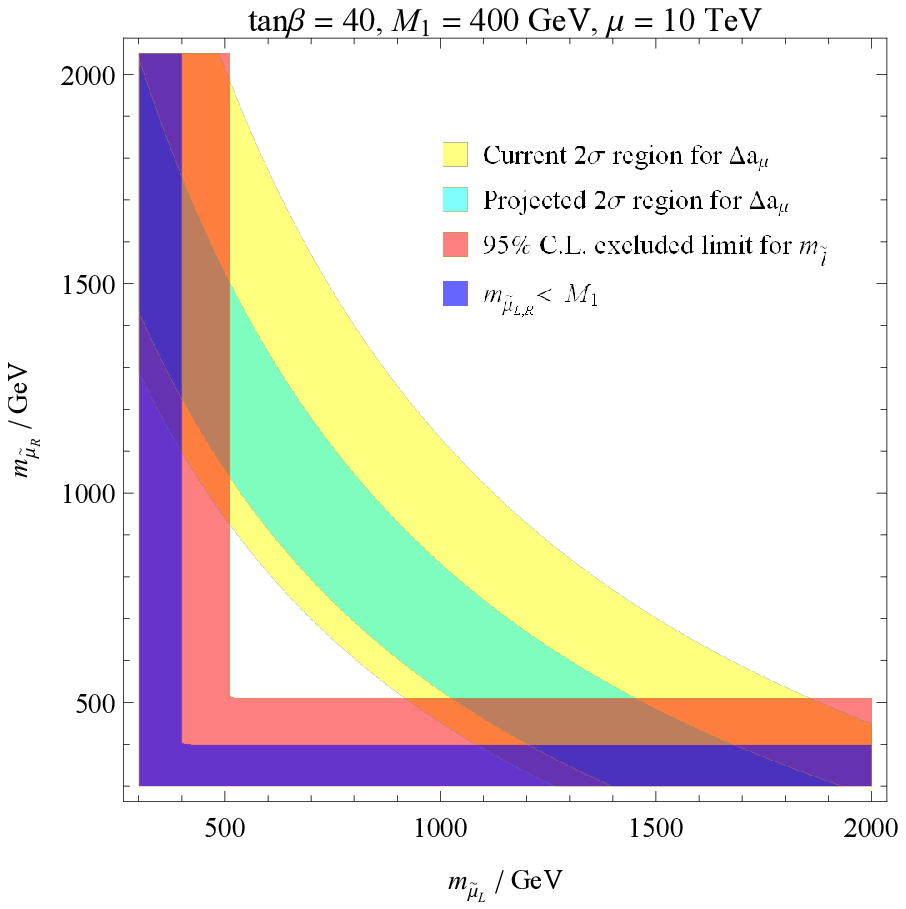}
\caption{
\label{fig8}
The phenomenological constraints in the plane of 
$(m_{\tilde{\mu}_L},m_{\tilde{\mu}_R})$.
The input parameters are fixed as
$\tan\beta = 40$,
$M_1 = 400$ GeV,
$\mu = 10$ TeV.
}
\end{figure}

For a given value of the bino mass $M_1$, the smuon masses are subjected to two constraints: (i) the requirement that the smuons must be heavier than the lightest bino-like neutralino $\tilde{\chi}^0_1$, (ii) the LHC lower limit (\ref{8b2}) on $m_{\tilde{l}}$.
We can see from both Figures \ref{fig6} and \ref{fig7} that when $M_1 < 510$ GeV, the constraint (ii) is more severe.
When $M_1 > 510$ GeV, the constraint (i) become more severe.
As an example, the allowed region in the plane of $(m_{\tilde{\mu}_L}, m_{\tilde{\mu}_R})$ is depicted in Figure \ref{fig8} for the case of 
$\tan\beta = 40$,
$M_1 = 400$ GeV,
$\mu = 10$ TeV.
The choice of $M_1$ here satisfies the constraint from the LHC searches, but smaller than the lower bound for sleptons.
Therefore, the 95\% C.L. excluded limit for the slepton is more severe than the constraint $m_{\tilde{\mu}} < M_1$.
In this case, the current upper bound for the smuon masses can be derived from the figure as 
$m_{\tilde{\mu}_L} \lesssim 1850$ GeV,
and 
$m_{\tilde{\mu}_R} \lesssim 2000$ GeV.
These upper bounds will be significantly reduced in the near future with the projected E989 experiment results (\ref{da_projected}):
\begin{eqnarray}
m_{\tilde{\mu}_L}  \lesssim 1450 \text{ (GeV)},	\\
m_{\tilde{\mu}_R}  \lesssim 1500 \text{ (GeV)}. 
\end{eqnarray}
Such parameter regions with the light neutralino and the light smuons will be accessible in the near future at the LHC Run 3 and the High Luminosity LHC (HL-LHC).

\section{Conclusion}

The new results of the LHC search for neutralino, chargino and slepton productions in final states with two or three leptons at $\sqrt{s}=13$ TeV with the integrated luminosity of 36.1 fb$^{-1}$ show the improved lower limits on their masses.
On the other hand, the recent study on the muon anomalous magnetic moment 
\cite{Keshavarzi:2018mgv} pointed out that the discrepancy between the SM prediction and the measured values can raise from 3.7$\sigma$ up to 7.0$\sigma$ with the projected E989 experiment assuming the unchanged center value of this quantity.
From an optimistic point of view, it is likely that we are close to finding something related to the muon $g-2$.
In this paper, we have investigated the GUT-inspired MSSM with the universal gaugino masses at the GUT scale.
In order to explain the SM-like Higgs boson of 125 GeV, the squarks are heavy and decoupled from physics at the TeV scale.
The light superparticles include the neutralinos ($\tilde{\chi}^0_{1,2}$), the charginos ($\tilde{\chi}^\pm_1$) and the sleptons of the first two generations.
Beside the muon $g-2$ constraints, we have taken into account the recent LHC bounds for their masses.
For large values of $\mu$ and $\tan\beta$, the vacuum metastability condition becomes important.
The allowed windows for each input parameters have been pinned down. Especially, we obtain the upper bounds for the bino mass $M_1$ and the smuon masses from these constraints.
It is pointed out that the projected muon $g-2$ measurement will play a crucial role in reducing these upper bounds.
Since these windows are applied for a few parameters corresponding to the light SUSY sector at low energies, the results can be applied to various SUSY GUT models with universal gaugino masses at the GUT scale.
The small mass regions of the allowed parameter space will be accessible in the near future at the LHC Run 3 and the HL-LHC.

With the improvement in the E989 experiment, the uncertainty on the deviation (\ref{da_projected}) between the SM prediction and the measured value of the muon $g-2$ 
that controls the allowed ranges of input parameters 
is dominated by the uncertainty of the SM prediction (\ref{SMprediction}).
Therefore, a reduction of the SM estimation uncertainty in the future will result in more severe limits on these parameters.
On the other hand, due to missing higher order diagrams in the SUSY calculation, the additional theoretical uncertainty relating to their contributions \cite{Stockinger:2006zn, Athron:2015rva} 
may enlarge the allowed parameter space.
Considering this uncertainty, the full two-loop calculation of the SUSY contribution to the muon $g-2$ is essential to increase the accuracy of the analysis and to obtain more stringent constraints on the input parameters.

\section*{Acknowledgment}

H.M.T. would like to thank Duc Minh Tran for reading the manuscript and useful comments.
This research is funded by Vietnam National Foundation for Science and Technology Development (NAFOSTED) under grant number 103.01-2017.301.


\end{document}